\documentclass[showpacs,prl,twocolumn,superscriptaddress]{revtex4}

\usepackage{graphicx}

\newcommand{\be}{\begin{equation}}
\newcommand{\ee}{\end{equation}}
\newcommand{\ba}{\begin{eqnarray}}
\newcommand{\ea}{\end{eqnarray}}
\newcommand{\bd}{\begin{description}}
\newcommand{\ed}{\end{description}}

\renewcommand{\iota}{{\bf 1}}
\def\rellow#1#2{Mathrel{Mathop{\kern 0pt #1}\limits_{#2}}}

\begin{document}

\title{Delayed Scattering of Solitary Waves from Interfaces in a Granular Container}
\author{Lautaro Vergara}
\email{lvergara@lauca.usach.cl}
\affiliation{Departamento de F\'{\i}sica, Universidad de Santiago de Chile, Casilla 307,
Santiago 2, Chile }
\date{today}

\begin{abstract}
In granular media, the characterization of the behavior of
solitary waves around interfaces is of importance in order to look
for more applications of these systems. We study the behavior of
solitary waves at both interfaces of a symmetric granular
container, a class of systems that has received recent attention
because it posses the feature of energy trapping. Hertzian contact
is assumed. We have found that the scattering process is elastic
at one interface, while at the other interface it is observed that
the transmitted solitary wave has stopped its movement during a
time that gets longer when the ratio between masses at the
interfaces increases. The origin of this effect can be traced back
to the phenomenon of gaps opening, recently observed
experimentally.

\end{abstract}

 \maketitle

%\section{Introduction}

The propagation of a perturbation in a chain of beads in Hertzian
contact possesses soliton-like features, as first observed by
Nesterenko \cite{Nesterenko}. Several studies, experimental
\cite{Nesterenko1,Coste} as well theoretical
\cite{Wattis,Nesterenko2} have confirmed the existence of such
soliton-like pulses. Despite the large amount of recent work on
the subject
\cite{Sen1,Hinch,Chatterjee,Manciu,SenM,SenMan,Coste,Hong,SenProc,Naka,Rosas,Sen2,Nesterenko2},
the physics of granular media remains a challenge and new effects
are there to be discovered and studied. Enlarging the number of
(engineering) applications of such new effects needs a complete
understanding of the dynamics of such granular media.

The simplest granular systems are one-dimensional chains of
elastic spheres. If they are in Hertzian contact, the spheres may
be considered as point masses interacting through massless
nonlinear springs with elastic force $F=k\delta ^{3/2}$, where
$\delta $ is the overlap of contacts and $k$ is the spring
constant (a function of the material properties)
\cite{Nesterenko}. Let $x_{i}(t)$ represents the displacement of
the center of the
$i$-th sphere from its initial equilibrium position, and assume that the $i$%
-th sphere, of mass $m_{i}$, has neighbors of different radii
(and/or mechanical properties). Then, in absence of load and in a
frictionless medium, the equation of motion for the $i$-th sphere
reads
\begin{equation}
m_{i}\frac{d^{2}x_{i}}{dt^{2}}%
=k_{1}(x_{i-1}-x_{i})^{3/2}-k_{2}(x_{i}-x_{i+1})^{3/2},  \label{uno}
\end{equation}%
where it is understood that the brackets take the argument value if they are
positive and zero otherwise, ensuring that the spheres interact only when in
contact.

The interaction of a solitary wave with the boundary of two "sonic
vacua" (meaning that the system does not support linear sound
waves if not precompressed) was studied for the first time
experimentally as well numerically in \cite{Nesterenko2} (see also
\cite{Nest,Nest3,Melo,Vergara} for a recent study).

In this work we make a detailed numerical study of the propagation
of solitary waves in a linear chain of beads composed of three
"sonic vacua", as shown in Fig. 1, that is, a granular container
\cite{HA,H}. This kind of systems are of interest because in them
one can find the phenomenon of energy trapping. It will be assumed
that all spheres have the same mechanical properties and that both
ends of the chain are free to move. We have found that the
scattering process is elastic at one interface, while at the other
interface it is observed that the transmitted solitary waves take
a long time to be released from one of the interfaces; this time
gets longer when the ratio between masses at the interfaces
increases. The origin of this effect can be traced back to the
phenomenon of gaps opening, recently observed experimentally
\cite{Nest,Nest3}. As far as we know, the effect found here has
not been yet observed in experiments.

Consider a set of spheres with two different radii $a$ and $b$. It
is known
that adjacent spheres of radii $a$ and $b$ will interact with a force $%
F=k_{ab}\delta^{3/2}$, where
\begin{equation}
k_{ab}=\frac{\sqrt{ab/(a+b)}}{2\theta },  \label{dos}
\end{equation}
with
\begin{equation}
\theta =\frac{3(1-\nu ^{2})}{4E}  \label{tres}
\end{equation}
and $E$ is the Young modulus and $\nu $ the Poisson ratio of the
bead material.

%********|*********|*********|*********|*********|*********|*********|****
%********|*********| Fig1:           |*********|*********|****
%********|*********|*********|*********|*********|*********|*********|****
\begin{figure}[tbp]
\centering %\vspace{-.5cm}
\hspace{-.4 cm} \includegraphics[width=.40\textwidth]{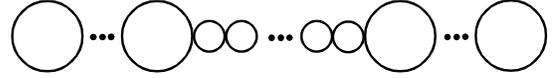} \vspace{%
-.1cm} \caption{{Schematic granular container used in the
calculations.}} \label{panel}
\end{figure}
%********|*********|*********|*********|*********|*********|*********|****

We will consider the scattering of solitary waves in a system like
the one of Fig. 1, consisting of a total of $M$ beads. There are
two set of beads with $N_{1}$ beads located on the lhs, $N_{2}$ on
the rhs, both sets have beads with radii $a$ and masses $m_{1}$.
Between them there are $L$ beads with radii $b$ ($a>b$) and masses
$m_{2}$. Beads displacements are governed by a set of equations of
motion that can be readily obtained from the successive
application of Eq. (\ref{uno}), having in mind that the equation
of motion for the first (resp. the last) sphere only includes the
second (resp. the first) term, in case when there is no wall (as
we here assume). Spring constants are $k_{bb}$ in the middle,
$k_{aa}$ at right and left hand sides and $k_{ab}$ at the
interfaces of the granular system.

In order to have realistic results, we shall assume that the
system consists of stainless-steel beads (see \cite{Coste} for
their properties), with radii $a=4$ mm and $b=2$ mm. The number of
beads is $N_{1}=30$, $N_{2}=20$ and $L=200$. We also choose
$\beta=10^{-5}$ m, $2.36 \,\times 10^{-5}$ kg and $\alpha = 1.0102
\,\times 10^{-3}$ s as units of distance, mass and time,
respectively. Through out the paper we assume that initially all
beads are at rest, except for the first bead at the left side of
the chain. This bead is supposed to have a nonzero value of
velocity in order to generate the soliton-like perturbation in the
chain. We shall choose the following initial conditions
\begin{eqnarray*}
u_{i}(0) &=&0,\text{ \ }i=1,\ldots ,M,\text{ \ }\dot{u}_{1}(0)=101.02 \,\beta/\alpha, \\
\dot{u}_{i}(0) &=&0,\text{ \ }i=2,\ldots ,M.  \label{init}
\end{eqnarray*}
This initial impact velocity corresponds to 1 m/s and therefore it
is in the regime where plastic deformation can be neglected. The
system is studied numerically by using an explicit Runge-Kutta
method of 5th order based on the Dormand-Prince coefficients, with
local extrapolation. As step size controller we have used the
proportional-integral step control, which gives a smooth step size
sequence.

%********|*********|*********|*********|*********|*********|*********|****
%********|*********| Fig2:           |*********|*********|****
%********|*********|*********|*********|*********|*********|*********|****
\begin{figure}[h]
\centering %\vspace{-.5cm}
\hspace{-.5 cm} \includegraphics[width=.34\textwidth]{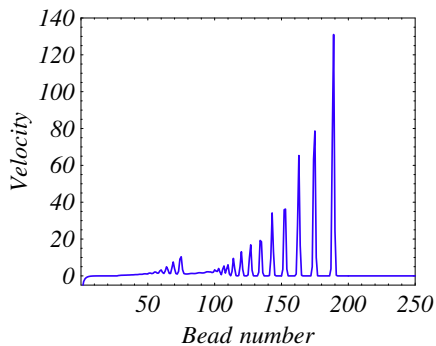} \vspace{%
-.3cm} \caption{{Velocity of beads (in program units) as a
function of bead number. Primary and secondary multipulse
structures.}} \label{panel}
\end{figure}
%********|*********|*********|*********|*********|*********|*********|****

As the solitary wave gets the interface a multipulse structure is
generated but no backscattered solitary wave is observed. This
last phenomenon has been explained by Nesterenko et al.
\cite{Nest} as due to the opening of gaps in the vicinity of left
interface. These effects originate from the discreteness of
inertia and the nonlinearity of the interaction; they were first
observed by Nesterenko and coworkers \cite{Nesterenko2}.

As the multipulse structure moves into the light system, there
remains some energy behind the interface and after a while a
second multipulse structure emerges, with similar characteristics
than the first one but with less energy \cite{Vergara}. This is
shown in Fig. 2. (It can be shown that the opening of gaps in the
vicinity of the left interface is also responsible for the
emergence of this structure).

%********|*********|*********|*********|*********|*********|*********|****

%********|*********|*********|*********|*********|*********|*********|****
%********|*********| Fig3:           |*********|*********|****
%********|*********|*********|*********|*********|*********|*********|****
\begin{figure}[h]
\centering %\vspace{-.5cm}
\hspace{-.5 cm} \includegraphics[width=.34\textwidth]{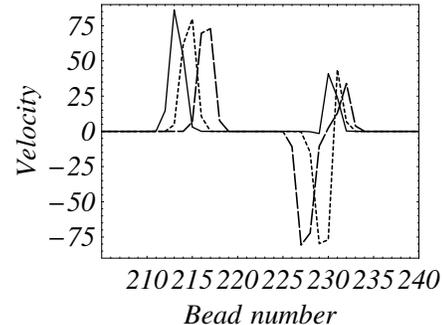} \vspace{%
-.2cm} \caption{{Velocity of beads (in program units) as a
function of bead number. Scattering of the leading pulse of the
multipulse structure at the second interface, for different times:
$t=1.303$ $\alpha$ (full line), $t=1.31$ $\alpha$ (dashed line)
and $t=1.32$ $\alpha$ (long-dashed line).}} \label{panel}
\end{figure}
%********|*********|*********|*********|*********|*********|*********|****

When the first multipulse structure interacts with the second
interface it gets immediately scattered and both transmitted and
reflected train of pulses appear in the right heavy system and in
the light system, respectively. Part of this process is shown in
Fig. 3.

Now let us see what happens when the backscattered pulses interact
with the left boundary of the granular potential well (see Fig.
4). It is observed that at time $t=2.32$ $\alpha$ the leading
backscattered pulse arrives at the interface. Contrary to what
happens at the right interface, the transmitted pulse moves
slightly forward till time $t=2.343$ $\alpha$ where it "freezes".
Even more intriguing is the fact that only beyond time $t=2.462$
$\alpha$ the transmitted pulse starts to move into the heavy
system. In our original units of time this means that the
transmitted pulse has stopped its movement during $1.2 \,\times
10^{-4}$ s approximately. This coincides with the fact that the
second pulse of the multipulse structure approaches close to the
interface. It is interesting to notice also that the backscattered
pulses scatter without delay at this interface.

%********|*********|*********|*********|*********|*********|*********|****
%********|*********| Fig4:           |*********|*********|****
%********|*********|*********|*********|*********|*********|*********|****
\begin{figure}[h]
\centering %\vspace{-.5cm}
\hspace{-0.5 cm} \includegraphics[width=.34\textwidth]{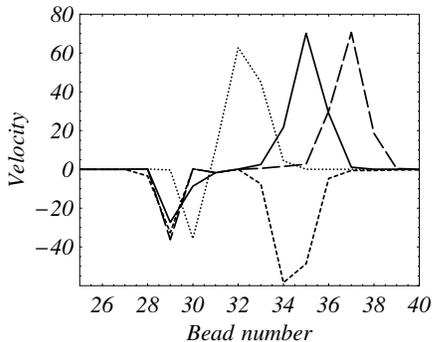} \vspace{%
-.2cm} \caption{{Velocity of beads (in program units) as a
function of bead number. Scattering of the first pulse of the
backscattered multipulse structure at the first interface, for
different times: $t=2.335$ $\alpha$ (dotted line), $t=2.35$
$\alpha$ (full line), $t=2.36$ $\alpha$ (long-dashed line) and
$t=2.462$ $\alpha$ (short-dashed line).}} \label{panel}
\end{figure}
%********|*********|*********|*********|*********|*********|*********|****

If we allow for the first bead to have more energy at $t=0$ s, it
is observed that the characteristics of the scattering process
around the second interface is similar to the one observed before:
the scattering is elastic. At the left interface the situation has
not essentially changed; there still is a delayed transmitted
pulse.

To get more insight into this scattering process, let us observe
the behavior of the velocity of beads as a function of time. We
shall analyze that behavior at both interfaces to see the
differences and try to find an explanation for this scattering. To
that end we shall fix our attention on those beads around the
interfaces.

%********|*********|*********|*********|*********|*********|*********|****
%********|*********| Fig5:           |*********|*********|****
%********|*********|*********|*********|*********|*********|*********|****
\begin{figure}[h]
\centering %\vspace{-.5cm}
\hspace{-1.29 cm} \includegraphics[width=.34\textwidth]{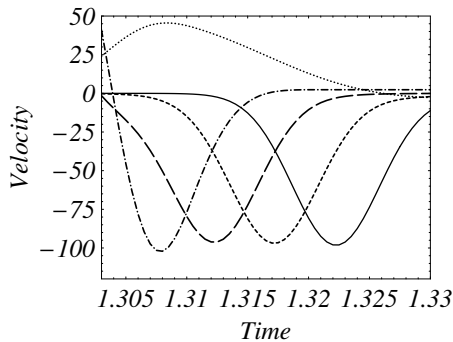} \vspace{%
-.2cm} \caption{{Velocity of beads (in program units) as a
function of time (in units of $\alpha$) for beads $227$ (full
line), $228$ (short-dashed line), $229$ (long-dashed line), $230$
(dashed-dotted line) and $231$ (dotted line).}} \label{panel}
\end{figure}
%********|*********|*********|*********|*********|*********|*********|****

In Figure 5 we show the velocity of beads between beads 227 and
231 (i.e., around the right interface) in the time interval 1.303
$\alpha$ and 1.33 $\alpha$. It can be straightforwardly
demonstrated that this behavior is analogous to that found in a
system with only two "sonic vacua", in case a solitary wave,
travelling in an unperturbed medium from the light to the heavy
system, scatters from the interface. So it it important to stress
that (in the case at hand) in the light system we have solitary
waves travelling in it; they correspond (at least) to those
leading pulses of the multipulse structure.

%********|*********|*********|*********|*********|*********|*********|****
%********|*********| Fig7:           |*********|*********|****
%********|*********|*********|*********|*********|*********|*********|****
\begin{figure}[tbp]
\centering %\vspace{-.5cm}
\hspace{-1.29 cm} \includegraphics[width=.34\textwidth]{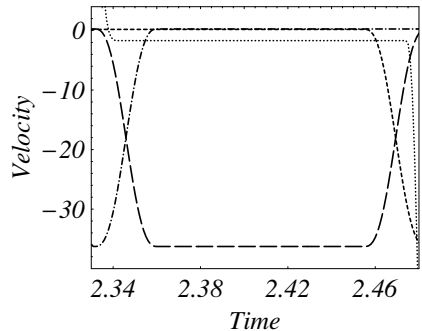} \vspace{%
-.2cm} \caption{{Velocity of beads (in program units) as a
function of time (in units of $\alpha)$ for beads $27$ (full
line), $28$ (short-dashed line), $29$ (long-dashed line) and $30$
(dotted line).}} \label{panel}
\end{figure}
%********|*********|*********|*********|*********|*********|*********|****

In Figure 6 the velocity of beads 27 to 30 (i.e., around the left
interface) is shown. It is notorious the big difference between
this behavior and the one found at the right interface. From here
we deduce that the origin of the delayed behavior in the
scattering process resides in the fact that, contrary to what
happens at the right interface, around the left interface beads
acquire a constant velocity in the interval $t\in (2.33 \alpha,
2.48 \alpha)$, where we have previously seen that the transmitted
pulse has stopped its movement during a long time. It is also
interesting to notice that bead 27 remains at rest during this
interval.

When changing the ratio of masses (in this case, the ratio $a:b$)
one observes that the interval that takes the transmitted pulse to
leave the interface increases. For example, one sees that by
keeping the mass and initial velocity of the impacting bead as
before, the intervals from the arrival to the left interface of
the first backscattered pulse to the instant when the transmitted
pulse starts to leave the interface are approximately $0.744$ ms,
for $a=6$ mm and $b = 2$ mm, and $1.244$ ms, for $a=8$ mm and $b =
2$ mm, respectively (compare with an interval of approximately
$0.143$ ms for the case $a=4$ mm and $b = 2$ mm. A numerical
experiment with $L=50$ beads in the interior of the container
shows that this time interval does not depend on $L$).

%********|*********|*********|*********|*********|*********|*********|****
%********|*********| Fig7:           |*********|*********|****
%********|*********|*********|*********|*********|*********|*********|****
\begin{figure}[h]
\centering %\vspace{-.5cm}
\hspace{-1.29 cm} \includegraphics[width=.34\textwidth]{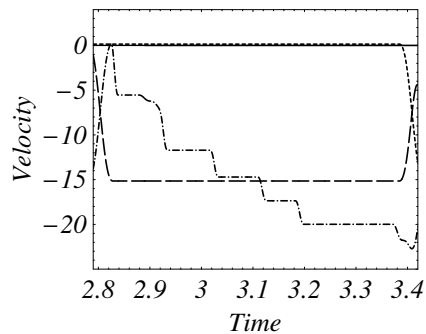} \vspace{%
-.2cm} \caption{{Velocity of beads (in program units) as a
function of time (in units of $\alpha$) for beads $27$ (full
line), $28$ (short-dotted line), $29$ (long-dashed line) and $30$
(dashed-dotted line).}} \label{panel}
\end{figure}
%********|*********|*********|*********|*********|*********|*********|****

In case $a=6$ mm and $b = 2$ mm, Fig. 7 shows the behavior of the
velocity of beads for beads 27 to 30 in the interval $t\in (2.8,
3.392)$ $\alpha$, where the transmitted pulse remains "frozen" at
the interface. Notice that such behavior for beads 27, 28 and 29
is essentially the same as the one observed in Fig. 6 for the same
beads (in particular, bead 27 remains at rest). This confirms that
the reason for the behavior observed in this work resides in the
fact that some beads near the interface acquire a constant
velocity during the interval where the transmitted solitary wave
retards its movement. Now, a constant velocity means that no
forces are acting on them and therefore the phenomenon of gaps
opening occurs.

Using a detailed numerical approach, we have studied the
scattering of solitary waves in a granular container consisting of
three "sonic vacua" with Hertzian contact. We have found that the
scattering process is elastic at the second interface, while at
the first interface it is observed that the transmitted solitary
wave has stopped its movement during a time that gets longer when
the ratio between masses at the interfaces increases. At the same
time, the reflected pulses appear to scatter elastically from the
first interface. The opening of gaps in the vicinity of the left
interface plays a crucial role in the observed behavior. The
understanding of this kind of behavior may be of help for
applications of energy-trapping granular containers.

%%%%%%%%%%%%%%%%%%%%%%%%%%%%%%%%%%%%%%%%%%%%%%%%%%%%%%%%%%%%%%%%%%

%\subsection*{Acknowledgements}

I thank Prof. Vitali F. Nesterenko for useful comments,
suggestions and for pointing my attention to the phenomenon of
gaps opening and Refs. \cite{HA,H}. I acknowledge comments from
Dr. Ra\'ul Labb\'e. This work was partially supported by
DICYT-USACH No. 04-0531VC.
%%%%%%%%%%%%%%%%%%%%%%%%%%%%%%%%%%%%%%%%%%%%%%%%%%%%%%%%%%%%%%%%%%%%%


\begin{thebibliography}{99}
\bibitem{Nesterenko} V.F. Nesterenko, J. Appl. Mech. Tech. Phys.
\textbf{5}, 733 (1983).

\bibitem{Nesterenko1} A.N. Lazaridi and V.F. Nesterenko, J. of Appl. Mech. Tech. Phys.
\textbf{26}, 405 (1985).


\bibitem{Coste} C. Coste, E. Falcon, and S. Fauve, Phys Rev. E
\textbf{56}, 6104 (1997).

\bibitem{Wattis} G. Friesecke and J.A.D. Wattis, Commun. Math. Phys. \textbf{%
161}, 391 (1994).

\bibitem{Nesterenko2} \textit{Dynamics of Heterogeneous Materials} V.F.
Nesterenko (Springer-Verlag, New York, 2001); V.F. Nesterenko,
A.N. Lazaridi and E.B. Sibiryakov, Jour. Applied Mech. Tech.
Phys., \textbf{36}, 166 (1995).

\bibitem{Sen1} S. Sen and R.S. Sinkovits, Phys. Rev. E \textbf{54},
6857 (1996).

\bibitem{SenM} S. Sen, M. Manciu and J.D Wright, Phys. Rev. E
\textbf{57}, 2386 (1998).

\bibitem{Hinch} E.J. Hinch and S. Saint-Jean, Proc. R. Soc. London, Ser. A
\textbf{455}, 3201 (1999).


\bibitem{Chatterjee} A. Chatterjee, Phys. Rev. E \textbf{59}, 5912
(1999).

\bibitem{Hong} J. Hong and A. Xu, Phys. Rev. E \textbf{63}, 061310 (2001).

\bibitem{Manciu} M. Manciu, S. Sen and A.J. Hurd, Physica (Amsterdam)
\textbf{157D}, 226 (2001).

\bibitem{SenMan} S. Sen and M. Manciu, Phys. Rev. E \textbf{64},
056605 (2001).

\bibitem{SenProc} S. Sen \textit{et al.}, in \textit{Modern Challenges in
Statistical Mechanics: Patterns, Noise and the Interplay of
Nonlinearity and Clomplexity}, edited by V. M. Kenkre and K.
Lindenberg, AIP Conf. Proc. No. 658 (AIP, New York, 2003), p. 357.

\bibitem{Naka} M. Nakagawa, J. H. Agui, D. T. Wu, and D. V. Extramiana,
Granular Matter \textbf{4}, 167 (2003).

\bibitem{Rosas} A. Rosas and K. Lindenberg, Phys. Rev. E
\textbf{69}, 037601 (2004).

\bibitem{Sen2} \textit{The Granular State}, edited by S. Sen and M.L. Hunt,
MRS Symposia Proceedings No. 627 (Material Research Society,
Pittsburg, 2001).

\bibitem{Nest}  V.F. Nesterenko, C. Daraio, E.B. Herbold and S. Jin, Phys. Rev. Lett. \textbf{95}, 158702
(2005).

\bibitem{Nest3}  C. Daraio, V.F. Nesterenko, E.B. Herbold and S. Jin, Phys. Rev. Lett. \textbf{96}, 058002
(2005).

\bibitem{Melo} S. Job, F. Melo, A. Sokolow and S. Sen, Phys. Rev. Lett. \textbf{94}, 178002
(2005).

\bibitem{Vergara} L. Vergara, Phys. Rev. Lett. \textbf{95}, 108002 (2005),
cond-mat/0503457.

\bibitem{HA} J. Hong and A. Xu, Appl. Phys. Lett., \textbf{81}, 4868
(2002).

\bibitem{H} J. Hong, Phys. Rev. Lett. \textbf{94}, 108001 (2005).



\end{thebibliography}
\end{document}